# STRUCTURE FLUCTUATIONS AND CONFORMATIONAL CHANGES IN PROTEIN BINDING

ANATOLY M. RUVINSKY

*Center for Bioinformatics, University of Kansas, 2030 Becker Drive*
*Lawrence, KS 66047, USA*
*ruvinsky@ku.edu*

TATSIANA KIRYS

*Center for Bioinformatics, University of Kansas, 2030 Becker Drive*
*Lawrence, KS 66047, USA*
*United Institute of Informatics Problems, National Academy of Sciences, Surganova 6*
*220012 Minsk, Belarus*

ALEXANDER V. TUZIKOV

*United Institute of Informatics Problems, National Academy of Sciences, Surganova 6*
*220012 Minsk, Belarus*

ILYA A. VAKSER

*Center for Bioinformatics and Department of Molecular Biosciences, University of Kansas, 2030 Becker Drive*
*Lawrence, KS 66047, USA*

Structure fluctuations and conformational changes accompany all biological processes involving macromolecules. The paper presents a classification of protein residues based on the normalized equilibrium fluctuations of the residue centers of mass in proteins and a statistical analysis of conformation changes in the side-chains upon binding. Normal mode analysis and an elastic network model were applied to a set of protein complexes to calculate the residue fluctuations and develop the residue classification. Comparison with a classification based on normalized B-factors suggests that the B-factors may underestimate protein flexibility in solvent. Our classification shows that protein loops and disordered fragments are enriched with highly fluctuating residues and depleted with weakly fluctuating residues. To calculate the dihedral angles distribution functions, the configuration space was divided into cells by a cubic grid. The effect of protein association on the distribution functions depends on the amino acid type and a grid step in the dihedral angles space. The changes in the dihedral angles increase from the near-backbone dihedral angle to the most distant one, for most residues. On average, one fifth of the interface residues change the rotamer state upon binding, whereas the rest of the interface residues undergo local readjustments within the same rotamer.

**1. Introduction**

Protein-protein interactions are a key component of life processes at the molecular level. Physical characterization of these interactions and their effects on protein structure are important for understanding mechanisms of protein function, our ability to manipulate protein function/structure and design effective drugs. Despite methodological advances and progress in computer technologies, accurate modeling of changes in protein structures upon binding is still one of the great challenges in modern bioinformatics and computational biology.[1] It is rooted in a rich variety of conformational changes that may occur in the unbound protein structures and upon binding. This includes structure fluctuations, allosteric changes, domain motions, local folding–unfolding transitions, transitions between regular secondary structure elements in "chameleon sequences," disorder-to-order transitions, and other changes in protein backbone and side chains. Several models were suggested to generalize the scope of the conformational changes observed: "lock-and-key model,"[2] the induced-fit model,[3] the allosteric model,[4] "the protein trinity model",[5] and the conformational selection and population shift.[6,7] The key questions of predictive modeling are what types of the changes hold upon binding and which of them make principal contributions. This study focuses on the structure fluctuation in the unbound proteins and conformation changes in the side chains upon binding. The focus on the side chains is motivated by the statistics of the unbound-to-bound changes in comprehensive non-redundant sets of protein complexes. The Benchmark set[8] has a $C^\alpha$ root mean square deviation (RMSD) < 2.2 Å for 84% of complexes. The DOCKGROUND set[9] used in this study has $C^\alpha$ RMSD ≤ 2 Å for 71% of the complexes. Thus, conformational changes in the side chains make the main contribution to structure changes upon binding for most complexes in the non-redundant sets.

In this paper, we present a classification of protein amino acids based on the normalized equilibrium fluctuations of the residue centers of mass. Normal mode analysis and an elastic network model were implemented in a program Vibe and applied to all unbound proteins from the DOCKGROUND set to calculate the fluctuations of each residue. The effect of structure packing on the fluctuations in solvent and crystal is discussed. Our classification is compared with a classification based on the normalized B-factors. We suggest that the use of the B-factors results in the underestimation of the Gly flexibility in solvent. We discuss relations between the scale of fluctuations and amino acid propensities for the secondary structure elements and protein disorder. The results showed that protein loops and disordered fragments are enriched with highly fluctuating residues and depleted with the weekly fluctuating ones. The analysis suggests a hypothesis that protein unfolding originates from highly fluctuating residues and their structural neighbors. Fluctuations of the interface residues were compared with other surface residues. Some residues were found to be more stable at the interface and form low-mobility patches. To calculate the distribution functions of the dihedral angles, the configuration space was divided into cells by a cubic grid. Correlation between bound and unbound interface/non-interface conformations, as well as between the interface and non-interface conformations in the bound/unbound states increases for larger grid step sizes in the dihedral angles space. The differences induced by binding forces disappear at the intervals comparable with the size of the clusters of the side-chain conformations mapping the rotamers in the configuration space. The average conformation changes in each of the dihedral angles are analyzed for all amino acid types. The scale of the changes depended on the side chain length and the proximity of the dihedral angle to the protein backbone. For most amino acids, the average conformational change increased from the near-backbone dihedral angle to the most distant one. The opposite trend was found in the residues with symmetric terminal groups (Phe, Tyr, Asp and Glu). Only one fifth of the interface residues changed its rotamer upon binding. The rest of the interface residues had local readjustments within the same rotamer.

## 2. Effects of protein sequence composition and structure packing on residue fluctuations

To investigate side-chain fluctuations, we designed an elastic network model with the nodes placed in the residue centers of mass and an inter-residue potential accounting for the distribution of interatomic interactions between the residues (the packing density). The approach is implemented in a program Vibe for the coarse-grained normal mode analysis in proteins.[10] Vibe was applied to 184 unbound proteins from 92 non-obligate complexes selected from the DOCKGROUND non-redundant set. A mobility ratio was calculated for each residue as a normalized mean square fluctuation in a protein: $R = \langle \vec{r}_k^2 \rangle / \langle r_{av}^2 \rangle$, where $\langle \vec{r}_k^2 \rangle$ is a mean square deviation of the center of mass of the $k$th residue from its equilibrium position due to thermal vibrations, angular brackets denote a Boltzmann average over the normal modes,[10] $r_{av}^2 = \sum_{k=1}^{N} \langle \vec{r}_k^2 \rangle / N$ is the average fluctuation of all the residue centers of mass in a protein with the $N$ residues. Mobility ratios averaged over all proteins are summarized in Table 1 for each amino acid type. Ranking amino acids over the average mobility ratio yields a three-group classification: (I) highly fluctuating - Gly, Ala, Ser, Pro, Asp; the mobility ratio > 1, (II) moderately fluctuating - Thr, Glu, Asn, Lys, Cys, Gln, Arg, Val; 0.7 ≤ mobility ratio ≤ 1, and (III) weakly fluctuating - His, Leu, Met, Ile, Tyr, Phe, Trp; mobility ratio < 0.7.

Interestingly, the moderately and weakly fluctuating Groups II and III are different in the degree of the hydrophilicity. The Group II is largely polar/hydrophilic, whereas the Group III is mostly nonpolar/hydrophobic. This difference is a consequence of the non-homogeneous distribution of the amino acids in proteins. Polar residues prefer conformations that are more exposed to water. Nonpolar residues prefer buried conformations that minimize the solvent accessible surface area. On the surface, a residue has fewer nearest neighbors than in the core, and thus experiences greater fluctuations.

Group I of the highly fluctuating residues shows higher propensities for protein loops than for regular secondary structure elements (α-helices and β-strands). Since the residue mobility increases with the temperature increase, we can hypothesize that the nucleation of the unfolded phase starts from the protein loops and clusters of the highly fluctuating residues. It further suggests that local and global protein thermostability can be increased by mutations replacing highly fluctuating residues with the weakly fluctuating ones.

Highly mobile Group I residues participate in local structure fluctuations (folding-unfolding reactions) that can determine/control protein function and propagate external perturbations through protein structures.[11-14]

Table 1. Amino acid classification based on the mobility ratios.

| Amino acid | Average mobility ratio | Side-chain polarity | Secondary structure propensity[15] | Protein disorder propensity[5] (D/O - disorder/order-promoting, N – neither disorder nor order promoting) | Normalized B-factors[16] |
|---|---|---|---|---|---|
| Gly | 2.33 | nonpolar | loop | D | 1.03 |
| Ala | 1.51 | nonpolar | helix | D | 0.98 |
| Ser | 1.34 | polar | loop | D | 1.05 |
| Pro | 1.11 | nonpolar | loop | D | 1.05 |
| Asp | 1.04 | polar | loop | N | 1.07 |
| Thr | 0.99 | polar | β structure | N | 1.00 |
| Glu | 0.96 | polar | helix | D | 1.09 |
| Asn | 0.94 | polar | loop | O | 1.05 |
| Lys | 0.88 | polar | helix | D | 1.10 |
| Cys | 0.86 | nonpolar | β structure | O | 0.91 |
| Gln | 0.84 | polar | helix | D | 1.04 |
| Arg | 0.75 | polar | helix | D | 1.01 |
| Val | 0.74 | nonpolar | β structure | O | 0.93 |
| His | 0.69 | polar | loop | N | 0.95 |
| Leu | 0.67 | nonpolar | helix | O | 0.94 |
| Met | 0.67 | nonpolar | helix | N | 0.95 |
| Ile | 0.63 | nonpolar | β structure | O | 0.93 |
| Tyr | 0.44 | polar | β structure | O | 0.93 |
| Phe | 0.42 | nonpolar | β structure | O | 0.92 |
| Trp | 0.34 | nonpolar | β structure | O | 0.90 |

Comparing our classification with a classification suggested by Dunker et al[5] to characterize amino acid propensities for protein disorder (Table 1), one can notice a striking difference between Groups I and III. The majority of the Group I of the highly fluctuating residues belongs to the disorder-promoting residues that are often found in the disordered segments. The majority of the Group III of the weakly fluctuating residues belongs to the order-promoting residues that are less frequently found in the disordered segments. The relation between the two classifications is easily explained by the dynamic nature of the protein disorder. Protein segments enriched with the highly mobile residues are unable to fold into a unique structure because of the fluctuations that prevent its formation. If, nevertheless, ordered, such segments tend to form irregular secondary structure elements (loops), because large fluctuations also tend to destroy the long-range order.

The results (Table 1) show that Gly has the highest mobility in our classification. Although this agrees with common knowledge, scales based on the normalized average B-factors of the backbone atoms put Gly, "generally considered to be the most flexible amino acid,"[16] in the middle of the flexibility scale (see Table 1 for the results from Vihinen et al[16]). It is important to understand how this relates to the well-known fact of increased flexibility in the Gly-rich proteins (e.g., collagen) and fragments (loops and structural hinges). It is likely that the conformation flexibility, structure fluctuations and large-scale internal motions are damped in the crystal in comparison to that in solution. As a result, the B-factor – a measure of atom vibrations in the solid state – underestimates protein flexibility in solution. As the residues with the largest flexibility, Glycines are affected the most. It is supported by the Halle's model[17] showing that the crystal packing (more precisely, the number of noncovalent neighbors including atoms from adjacent crystal cells) determines the B-factors in protein structures and by Eyal et al[18] study showing that the B-factors of the residues with the crystal contacts are less than those of the fully exposed residues. These results are in agreement with the observation that the B-factors of the side-chain atoms are greater than the B-factors of the backbone atoms.[19] The more exposed side-chain atoms have less nearest neighbors than the largely buried backbone atoms and thus have greater B-factors. An increase of the B-factors values from the near-backbone atoms to the outer atoms was first reported for myoglobin[20, 21] and lysozyme,[22] and then confirmed for a large set of proteins.[23] The mobility ratios calculated here by Vibe follow the same trend by increasing from the protein core to the surface for all amino acid types.[10]

Another closely related evidence of the packing density effect is that the B-factors increase with the increase of the atom solvent accessible area.[19] A large-scale analysis of the catalytic residues in a set of 178 enzymes showed that 80% of these residues have a relative solvent accessibility < 20% of that in the fully exposed residues.[24] These buried residues have lower B-factors than the non-catalytic residues.[24,25] Enhanced packing can produce low-mobility interface pads that can be "landing ground" for binding proteins. These pads will be enriched with the weakly fluctuating residues of the Group III and thus are mostly non-polar. Besides the Group III residues, the low-mobility pads can also include residues from the Groups I and II if their interface fluctuations are substantially reduced. The most significant difference between interface and non-interface fluctuations relates to Gly, Ala, Ser, Cys, Leu and Trp (Fig. 1). Gly, Ala and Ser have the highest abilities to fluctuate, and Trp is always the most stable residue, on the non-interface surface and at the interface. At the interface, these six residues lose > 0.25 of their mobility ratio at the non-interface surface. For comparison, 0.3 and 0.35 are the ranges of mobility in Groups II and III (Table 1). On average, these residues show a 39% reduction in the mobility ratio. The reduction of the interface mobility relates mainly to fluctuations of the nonpolar residues with the exception of Ser, a polar residue. Attractions between nonpolar residues are considered a factor stabilizing protein complexes.[26] Most common residues at protein interfaces,[27,28] Gly, Ala, Leu, and Ser contribute to the continuous topology of the low-mobility pads. Cys and Trp are the most infrequent interface residues. The most conserved interface residues[29] Trp, Met and Phe demonstrate decreased mobility at the interface.

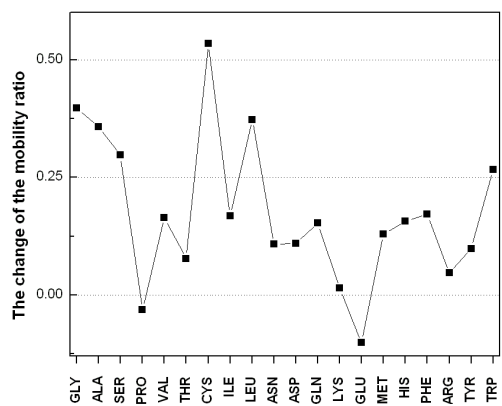

Figure 1. The difference between the non-interface surface and interface mobility ratios.

High similarity in packing densities of the backbone atoms of different residues[30] explains why the B-factors of these atoms are similar.[31] The normalized B-factors of the backbone atoms[16] (Table 1) vary by 18% of the value that is much smaller than the 85% variation of the mobility ratio of the residue center of mass. The larger variation of the mobility ratio suggests better sensitivity of this scale. It suggests the use the mobility ratios instead of normalized B-factors in the algorithms for predicting protein flexibility and rigidity from the sequence.

## 3. Side-chain configuration space in protein interactions

To reveal some of the general principles of unbound-to-bound conformational changes, we compared distribution functions of the interface and non-interface side chain dihedral angles in bound and unbound proteins in the DOCKGROUND set of the 233 non-obligate protein-protein complexes. The set contained the unbound structures of both proteins for 99 complexes and the unbound structure of one of the proteins for 134 complexes. The structures were selected from PDB, based on the following criteria: sequence identity between bound and unbound structures > 97%, and sequence identity between complexes < 30%, with homomultimers and crystal packing complexes excluded. To calculate the distribution function for the dihedral angles, the configurational space was divided into cells by a cubic *n*-dimensional grid (*n* is the number of dihedral angles in a side chain) and the corresponding probabilities of the interface and non-interface side-chain conformation in bound and unbound structures were calculated for each grid cell. An example of such distribution is shown in Figure 2. Figure 3 shows the Pearson correlation coefficient between bound and unbound interface (A) /non-interface (B) distribution

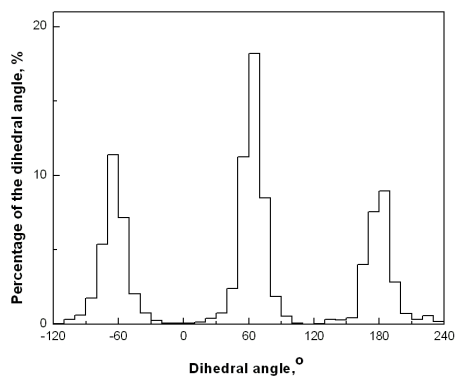

Figure 2. Distribution function of Ser dihedral angle.

functions calculated as

$$R = \frac{\sum_{i=1}^{n}(X_i - \bar{X})(Y_i - \bar{Y})}{\sqrt{\sum_{i=1}^{n}(X_i - \bar{X})^2} \sqrt{\sum_{i=1}^{n}(Y_i - \bar{Y})^2}} \quad (1)$$

for different grid steps in the dihedral angles space ($X_i$ and $Y_i$ are the probabilities of bound and unbound side-chain conformations in a grid cell $i$, $\bar{X}$ and $\bar{Y}$ are the average probabilities of bound and unbound side-chain conformations). To decrease the effect of the grid discretization, the grid was chosen randomly 100 times, and the average correlation coefficients were considered for each value of the grid step for the interface and non-interface side chains. Figure 3 shows that for the majority of the amino acid types, there is high correlation between bound and unbound probability functions. The exceptions ($R \leq 0.7$) are Met and Arg at the interface and the non-interface, Glu and Gln at the interface. In general, the correlation coefficients increase with the grid step increase. The interface correlation coefficients for Met and Arg are less than the corresponding non-interface correlation coefficients and reach 0.7 for the 70°/30° grid step used for the interface/non-interface statistics. This two-fold difference in the grid step suggests larger conformation changes in these residues at the interface.

The statistical analysis of the width of the peaks in the dihedral angle distribution functions shows that 70° is a typical width for all residues. The distance of 120° between two adjacent side-chain rotamers is larger than the rotamer width, and so libraries of bound and unbound rotamers should be similar (Kirys *et al*, in preparation). Analysis of the significance of the correlation coefficients was done for all amino acid types, at each grid step. The correlation coefficients are significantly different from zero (p<0.0001). Similar behavior was found for the correlations between interface and non-interface side-chain distribution functions in bound and unbound states. Figure 4 shows that low to moderate correlations exists at the 10° grid step for Gln, Lys, Glu, Met, His, Arg and Trp. All correlation coefficients exceed 0.7 at 50° grid step. The results suggest that the difference between bound and unbound conformations appears at small grid steps and disappear at large grid steps typically used in building rotamer libraries.

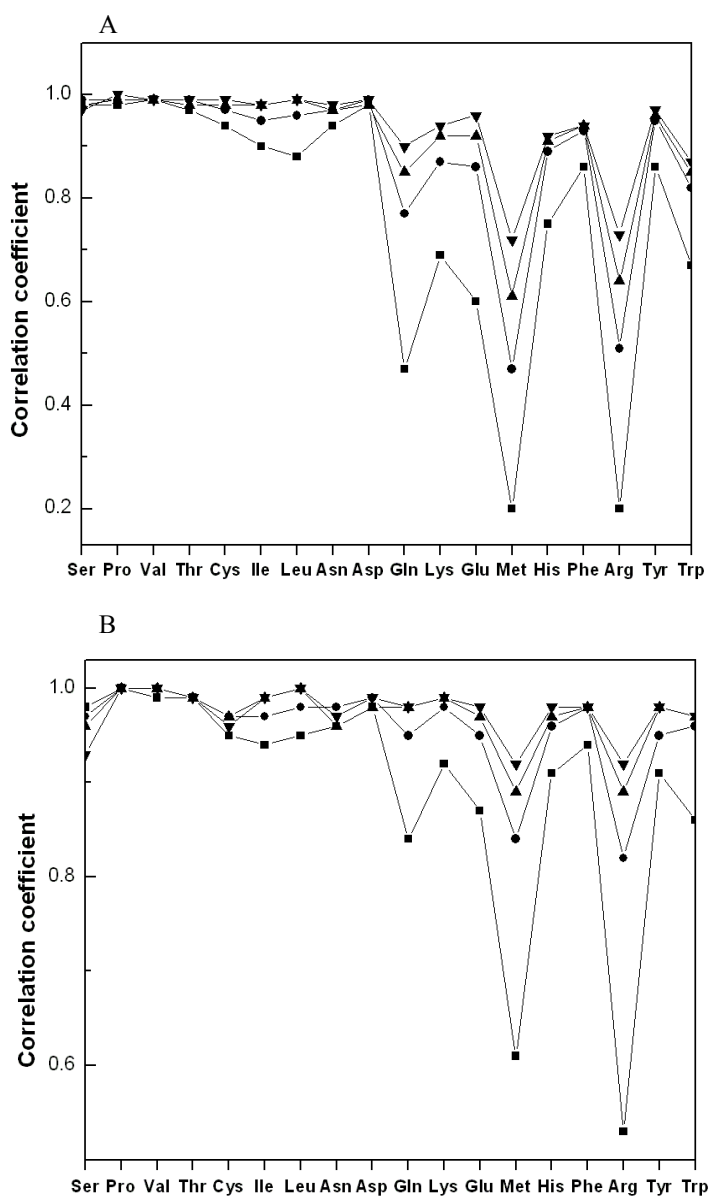

Figure 3. Correlation between dihedral angle distribution functions in bound and unbound proteins. (A) interface, (B) non-interface. The grid steps are 20 (■), 40 (●), 60 (▲) and 100° (▼).

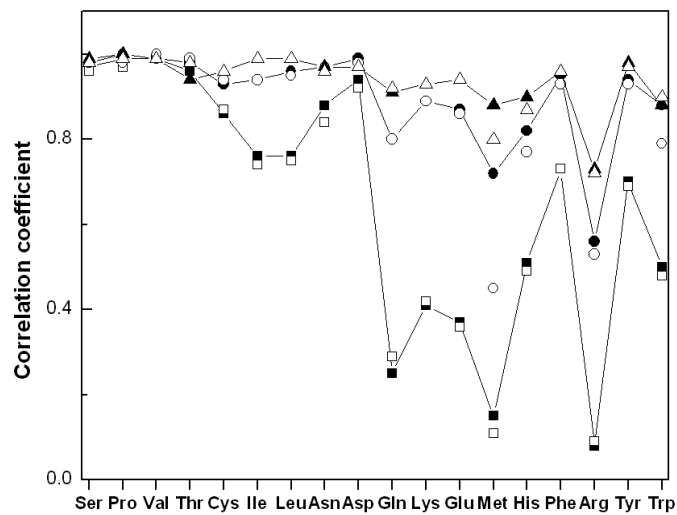

Figure 4. Correlation coefficient between the interface and non-interface distribution functions in the bound and unbound states. The grid steps are 10 (■,□), 30 (●,○) and 50° (▲,△). Black and white symbols correspond to bound and unbound states.

## 4. Unbound-to-bound conformational changes in the side chains

The results show that on average five interface side chains have a conformational transition > 100° upon protein-protein binding (Fig. 5). The most probable number of the transitions is 3 - 4. The average interface has 26 residues. If Ala, Pro and Gly are excluded, the average interface has 22 amino acids. Thus, only one fifth of the interface residues that are able to change the side-chain conformation do that upon binding. Thus, the induced fit mechanism applies to a small part of the interface. Most of the interface residues undergo local conformation changes (readjustments) not leading to the transitions between the rotamer states. Those local readjustments may or may not reflect specifics of protein-protein interactions. Thermal fluctuations of a side chain around its equilibrium rotamer populate a number of the intra-rotamer states. One of these states is selected upon binding. Such selection mechanism, followed by week changes of the internal energy and protein structures, can be named "lubricated lock-and-key model," meaning that the fluctuations provide fine-scale interface plasticity for the binding proteins. This model describes binding of 11% of the protein-protein complexes in the non-redundant DOCKGROUND set. On the other hand, significant energy is needed to induce an inter-rotamer transition in the side chains, to convert a low energy conformation into a high-energy one, or to rearrange protein domains (the induced fit mechanism). A combination of the induced fit with the lubricated lock-and-key model is a dominant binding mechanism according to the analysis of DOCKGROUND set.

Figure 6 shows the average changes of each dihedral angle in the interface amino acids. Most of the amino acids have larger changes of the outer angle in comparison with the near-backbone angle. An opposite trend, where the outer dihedral angle changes less than the near-backbone one, is observed in Phe, Tyr, Asp and Glu. It is explained by the reduced interval of variability of the outer dihedral angle due to the symmetry of the aromatic and charged terminal groups in these amino acids. All average changes are < $100^o$. This suggests that the most probable conformation change takes place within the same energy minimum and thus does not alter the side chain conformation, in agreement with the results in Figure 5. The angle changes increase with the increase in the number of the dihedral angles from 2 to 4. This relation between the number of the dihedral angles (the side-chain length) and the amplitude of the change may be explained by the difference in sampling times spent by short and long side chains to optimize their conformations upon binding. The longer residues establish interactions across interface earlier than the shorter residues and thus have more time for conformational sampling. The dihedral angle change decreases in residues with one dihedral angle (Ser, Val, Thr and Cys) with the increase of the amino acid's mass and the main moments of inertia. Week changes in the Cys residues may be also related to disulfide bonds imposing constraints on the side chain motions.

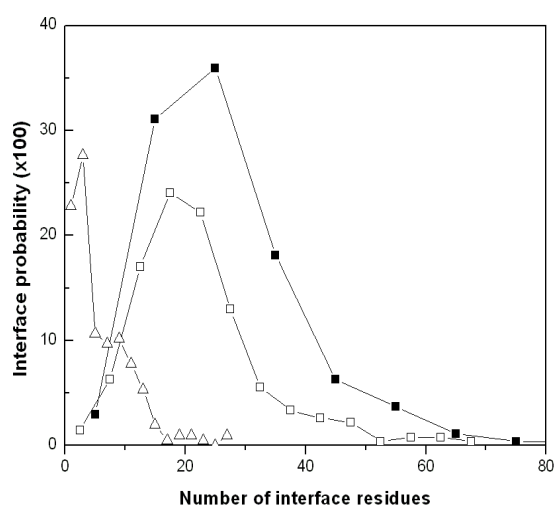

Figure 5. Probability of conformational transitions at the interface. The probability distribution of the number of transitions (△) as a function of the interface size. Probability distribution of the interface size: all residues included (■); Ala, Gly and Pro excluded (□).

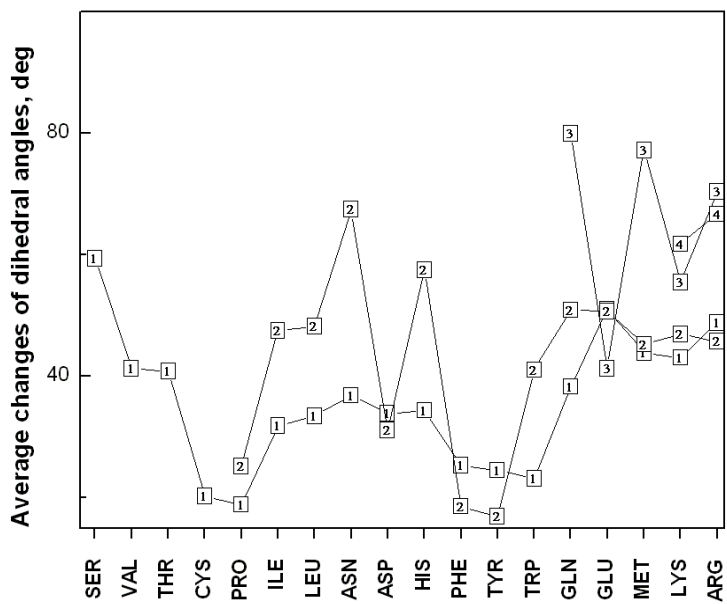

Figure 6. Average change of the dihedral angles in the side chains. The outlined numbers 1, 2, 3 and 4 correspond to the number of the dihedral angle in the side chains.

## 5. Conclusions

A classification of protein amino acids based on the normalized equilibrium fluctuations of the residue centers of mass was suggested. Normal mode analysis and an elastic network model were implemented in a program Vibe and applied to unbound proteins from the DOCKGROUND set to calculate the residue fluctuations. The results were compared with a flexibility scale based on the normalized B-factors. The analysis showed that use of the B-factors results in underestimation of Gly flexibility in solvent. Our classification correlated with the amino acid propensities for the secondary structure elements and protein disorder. The results showed that protein loops and disordered fragments are enriched with highly fluctuating residues and depleted with the weekly fluctuating ones. We suggested a hypothesis that protein unfolding proceeds from highly fluctuating residues and their structural neighbors. The abilities of interface and non-interface surface residues to fluctuate were compared. Some residues were shown to be more stable at the interface, forming continuous low-mobility patches.

To calculate the distribution function for the dihedral angles, the configuration space was divided into cells by a cubic grid. The results showed that the correlation between bound and unbound interface/non-interface side chain conformations, as well as between the interface and non-interface side chain conformations in the bound/unbound states increase with the grid step increase. The differences induced by binding forces disappear at the grid step comparable with the size of the clusters of the side-chain conformations.

The average conformation changes in each of the dihedral angles were analyzed for all amino acid types. The scale of the changes depended on the side chain length and the proximity of the dihedral angle to the protein backbone. In most amino acids, the average conformation change increased from the near-backbone dihedral angle to the most distant one. The opposite trend was found in the residues with symmetric terminal groups (Phe, Tyr, Asp and Glu). Only one fifth of the interface residues change its rotamer upon binding. The rest of the interface residues undergo local readjustments within the same rotamer.

**Acknowledgments**

The study was supported by NIH grant R01GM074255.